\documentstyle[preprint,epsfig,eqsecnum,aps]{revtex}
\tightenlines
\newcommand{\bea}{\begin{eqnarray}}
\newcommand{\eea}{\end{eqnarray}}
\newcommand{\beq}{\begin{equation}}
\newcommand{\eeq}{\end{equation}}
\begin{document}
\preprint{\parbox{4cm}{\flushright MZ-TH/95-32}}
\title{Spin-Momentum Correlations in Inclusive Semileptonic Decays of
Polarized $\Lambda_{\mathrm{b}}$-Baryons}
\author{J.G. K\"orner and D. Pirjol\footnote{Present address:
Floyd R. Newman Laboratory of Nuclear Studies, Cornell University,
Ithaca, New York 14853.}}
\address{Johannes Gutenberg-Universit\"at,
Institut f\"ur Physik (THEP), Staudingerweg 7,\\
D-55099 Mainz, Germany}
\date{\today}
\maketitle
\begin{abstract}
We consider spin-momentum correlations between the spin of the bottom
baryon $\Lambda_{\mathrm{b}}$ and the momenta of its decay products in its
inclusive semileptonic decay. We define several polar and azimuthal
spin-momentum correlation measures in different event coordinate systems.
The values of the spin-momentum correlation measures are calculated up to
${\cal O}(1/m_b^2)$ using the standard OPE und HQET methods. Some of the
measures turn out to be sufficiently large to make them good candidates
for a determination of the polarization of the $\Lambda_{\mathrm{b}}$ in
e.g. $\mathrm{Z}$ decays.
\end{abstract}
\narrowtext
\section{Introduction}
A few years ago the ALEPH Collaboration has measured the polarization of bottom
baryons $\Lambda_{\mathrm{b}}$'s originating from $\mathrm{Z}$ decays
\cite{PPE}. The ALEPH collaboration quoted a value for the polarization
$P=-0.23^{+0.24}_{-0.20}\pm 0.08$ which is significantly smaller than
what would be expected theoretically in the Standard Model ($P=-(0.60
\div 0.70)$) \cite{tr1}.
Recently a new measurement of the polarization has become available
from the OPAL collaboration \cite{OPAL}. They obtain the result
$P=-0.56^{+0.20}_{-0.13}\pm 0.09$ which is in accord with theoretical
expectations.

  The measurement of the ALEPH collaboration is based on the observation
of Bonvicini and Randall \cite{BR} that, with negatively polarized
$\Lambda_{\mathrm{b}}$'s, the spectra of the decay electrons and 
antineutrinos become harder and softer relative to unpolarized decay,
respectively, and that the fragmentation dependence of $\mathrm{b}\to
\Lambda_{\mathrm{b}}$ practically drops out in the ratio $y=\langle E_\ell
\rangle/\langle E_{\bar\nu}\rangle$. In a previous paper we have explored
possible improvements on such spectra related polarization measures
\cite{prev}. A promising candidate measure is, among others, the ratio
$y_2=\langle E_\ell^2\rangle/\langle E_{\bar\nu}^2\rangle$, a measurement
of which may help to reduce the errors in the original ALEPH analysis.

The method used by the OPAL collaboration \cite{OPAL} is to compare the
observed distribution of the ratio $E_\ell/E_\nu$ against a simulation
of this ratio using a JETSET Monte Carlo event generator. It is perhaps
worth mentioning that the distribution of this ratio is sensitive
to the precise shape of the $b\to\Lambda_b$ fragmentation function
\cite{prev}, which is not the case with the ratios $y_n=\langle E_\ell^n
\rangle/\langle E_{\bar\nu}^n\rangle$. A modified method was proposed in
\cite{prev} which avoids this problem, wherein the fragmentation dependence 
is eliminated between the two ratios $\langle E_\ell/ E_{\bar\nu}\rangle$ 
and $\langle E_{\bar\nu}/E_\ell\rangle$.

  In this paper we explore possibilities to determine the polarization
of the $\Lambda_{\mathrm{b}}$ through angular spin-momentum correlations
of the spin of the $\Lambda_{\mathrm{b}}$ and the momenta of its decay 
products in its inclusive semileptonic decays (the results of a 
preliminary  version of the present work have been presented in \cite{JGK}). 
We work in the rest frame of
the decaying $\Lambda_{\mathrm{b}}$ throughout and define various 
polarization measures which we compute up to ${\cal O}(1/m_b^2)$ in the
heavy mass expansion using the standard OPE and HQET approach to inclusive
semileptonic decays developed in \cite{CGG,Blok,MW}. The $O(\alpha_s)$ 
radiative corrections to some of these asymmetry parameters $(B_1, B_3)$
have been previously computed \cite{CJKK,CJ}.

  The analysis of the spin-momentum correlation measures in Sec.~2 makes
use of so-called helicity systems, in which the plane spanned by the three
final state momenta $\vec p_X,\,\vec p_\ell$ and $\vec p_\nu$ (event plane)
is in the ($x,z$) plane. The orientation of the polarization vector $\vec 
P$ is specified by two angles $(\vartheta,\,\varphi$) for which we compute
the angular decay distributions. In Sec.~3 we do the same exercise for the
so-called transversity systems, where the event plane defines the ($x,y$)
coordinate plane.

\section{Spin-momentum correlations in the helicity system}

 As we are analyzing the decay $\Lambda_{\mathrm{b}}\to X_c(p_X)+\ell^-
(p_\ell)+\bar\nu_\ell(p_\nu)$ in the rest frame of the
$\Lambda_{\mathrm{b}}$, the three momenta $\vec p_X,\,\vec p_\ell$ and
$\vec p_\nu$ lie in a plane -- the event plane. It is then a matter of
choice how to orient the event coordinate system relative to the event
plane and thereby relative to the polarization vector of the
$\Lambda_{\mathrm{b}}$. In this section we will discuss so-called
helicity systems in which the $z$-axis is in the event plane. It is then
convenient to define three coordinate systems according to the orientation
of the $z$-axis. Also one has to specify the orientation of the $x$-axis
for which one has two possible choices in each system. We thus define our
coordinate systems as
\bea
\mbox{system 1 : } \vec p_\ell\parallel z\quad &;&
\quad a:(\vec p_{\bar\nu})_x\geq 0
\qquad b:(\vec p_X)_x\geq 0\nonumber\\\label{1}
\mbox{system 2 : } \vec p_X\parallel z\quad &;&\quad a:(\vec p_\ell)_x\geq 0
\qquad b:(\vec p_{\bar\nu})_x\geq 0\\
\mbox{system 3 : } \vec p_{\bar\nu}\parallel z\quad &;&\quad a:(\vec p_X)_x
\geq 0 \qquad b:(\vec p_\ell)_x\geq 0\,.\nonumber
\eea
In this paper we shall always work in the systems 1a, 2a and 3a such that
$\vec p_{\bar\nu}, \vec p_{\ell}$ and $\vec p_X$, respectively, have positive
$x$-components. When using systems 1b, 2b and 3b the sign of the coefficient
$B$ in the angular decay distribution defined in Eq.(\ref{2}) below remains 
unchanged while the sign of the coefficient $C$ changes as can be seen by
making the transformation $\cos\varphi\to\cos(\varphi+\pi)=-\cos\varphi$.

It should be clear that the choice of the $z$-axis in the event plane is 
optional. Other possible choices would be to take the directions bisecting
any two of the three momenta directions $\vec p_X,\,\vec p_\ell,\,\vec
p_\nu$, etc. The above choice has been made for experimental convenience.

   In generic form the five-fold decay distribution (differential in $q_0,
\,q^2,\,\cos\theta,\,\cos\vartheta$ and $\varphi$) reads
\bea\label{2}
& &\frac{\mbox{d}\Gamma}{\mbox{d}q_0\mbox{d}q^2\mbox{d}\cos\theta\mbox{d}
\cos\vartheta\mbox{d}\varphi} =\\
& &\qquad\qquad\frac{1}{4\pi}
\Gamma_b\left\{\frac{\mbox{d}\hat\Gamma_A}{\mbox{d}\hat q_0
\mbox{d}\hat q^2\mbox{d}\cos\theta} +
P\left(\frac{\mbox{d}\hat\Gamma_B}{\mbox{d}\hat q_0\mbox{d}\hat q^2
\mbox{d}\cos\theta}\cos\vartheta +
\frac{\mbox{d}\hat\Gamma_C}{\mbox{d}\hat q_0\mbox{d}\hat q^2
\mbox{d}\cos\theta}\sin\vartheta\cos\varphi\right)\right\}\nonumber
\eea
where
\bea\label{3}
\Gamma_b = \frac{G_F^2|V_{cb}|^2m_b^5}{192\pi^3}
\eea
is the reference rate of the decay into three massless final particles.
Other symbols appearing in Eq.~(\ref{2}) are defined as follows.
The energy and the invariant mass squared of the virtual boson are denoted
by $q_0$ and $q^2$ respectively, with corresponding reduced quantities
$\hat q_0=q_0/m_b$ and $\hat q^2=q^2/m_b^2$.
The polar angle of the lepton $\ell^-$ in the ($\ell^-,\,\bar\nu_\ell$)
rest frame relative to the direction of $\vec p_X$ is denoted by $\theta$.
There is one unpolarized reduced rate function $\mbox{d}\hat\Gamma_A$ and
two polarized rate functions $\mbox{d}\hat\Gamma_B$ and $\mbox{d}\hat
\Gamma_C$. We shall sometimes also employ the notation
\bea
\frac{\mbox{d}\hat\Gamma_I}{\mbox{d}\hat q_0\mbox{d}\hat q^2\mbox{d}
\cos\theta} = I(\hat q_0,\hat q^2,\cos\theta)\,,\qquad I=A,B,C
\eea
with a corresponding notation for the once, twice and thrice integrated
forms.
The polar angle $\vartheta$ and the azimuthal angle $\varphi$ define the 
orientation of the polarization vector $\vec P$ in the helicity system
as drawn in Fig.1. Finally, $P=|\vec P|$ is the magnitude of the 
polarization of the $\Lambda_{\mathrm{b}}$.

  In Eq.~(\ref{2}) we have chosen the set of phase space variables ($\hat
q_0,\,\hat q^2,\,\cos\theta$). One could have equally well chosen the set
($\hat q_0,\,\hat q^2,\,y=2E_\ell/m_b$) where $E_\ell$ denotes the energy
of the lepton. Using the relation
\bea
y = -\hat p\cos\theta + \hat q_0
\eea
one has
\bea
\frac{\mbox{d}\hat\Gamma}{\mbox{d}\hat q_0\mbox{d}\hat q^2\mbox{d}y} =
-\hat p\frac{\mbox{d}\hat\Gamma}{\mbox{d}\hat q_0\mbox{d}\hat q^2\mbox{d}
\cos\theta} 
\eea
where $\hat p=\sqrt{\hat q_0^2-\hat q^2}$. However, the choice of variables
($\hat q_0,\,\hat q^2,\,\cos\theta$) has technical advantages when 
calculating the ${\cal O}(1/m_b^2)$ contributions to the rate expressions.
This can be seen as follows. The absorptive parts of the OPE expansion
give rise to higher order derivatives of the $\delta$-function, of the form
\bea\label{8}
\delta^{(n)}(\hat q_0 - \frac12(1-\rho+\hat q^2))
\eea
where $\rho=m_c^2/m_b^2$. When doing the $q_0$-integration the derivatives
of the $\delta$-function can be shifted to the integrand using partial 
integration, plus possible surface term contributions. When using the
($\hat q_0,\,\hat q^2,\,\cos\theta$)--set of phase-space variables, the
surface term contributions are identically zero \cite{Stephan}, whereas 
there are 
nonvanishing surface term contributions in the ($\hat q_0,\,\hat q^2,\,y$) 
phase space. 
In particular for $m_\ell\neq 0$ the surface term
contributions can become technically quite involved in the latter case and
lead to spurious singularities which have to be treated with care \cite{Gremm}.
Thus, when using the ($\hat q_0,\,\hat q^2,\,\cos\theta$) set of variables
the $\hat q_0$-integration can easily be done.

   Next we turn to the remaining $\hat q^2$-- and $\cos\theta$-integrations.
It turns out that the $\cos\theta$-dependence of the unpolarized rate 
function $A(\hat q^2,\,\cos\theta)$ and the polarized rate functions
$B(\hat q^2,\,\cos\theta),\,C(\hat q^2,\,\cos\theta)$ is particularly 
simple in system 2 and can easily be integrated. The $\cos\theta$-dependence
is so simple in this system since it is determined by bilinear forms of the
matrix elements of the Wigner $d^1_{mm'}(\theta)$ function. The requisite
$\cos\theta$-integrations can easily be done and one has \cite{Stephan}
\bea
\frac{\mbox{d}\hat\Gamma_A}{\mbox{d}\hat q^2} &=& 4\hat p(-2\hat q^4 +
\hat q^2(1+\rho) + (1-\rho)^2)(1-K_b)\\
\frac{\mbox{d}\hat\Gamma_B}{\mbox{d}\hat q^2} &=& 8\hat p^2
(1+\varepsilon_b)(2\hat q^2+\rho-1) \\
&-& K_b\left( 4\hat q^6 + \frac23 \hat q^4 - 6\hat q^4\rho + \frac83
\hat q^2(1-\rho) - 2(1-\rho)^3\right)\nonumber\\
\frac{\mbox{d}\hat\Gamma_C}{\mbox{d}\hat q^2} &=& 3\pi\hat p\sqrt{\hat q^2}
\left((1+\varepsilon_b)(\hat q^2+\rho-1) + \frac23K_b(\hat q^2-\rho+1)
\right)
\eea
where $\hat p = \frac12\sqrt{\hat q^4-2\hat q^2(1+\rho)+(1-\rho)^2}$,
$K_b$ is the mean kinetic energy of the heavy quark in the
$\Lambda_{\mathrm{b}}$ baryon and $\varepsilon_b$ is a spin dependent
forward matrix element on the $\Lambda_{\mathrm{b}}$. They are defined by \cite{MW}
\bea
K_b = -\frac{1}{2M_B}\langle\Lambda_{\mathrm{b}}(v)|\bar b\frac{(iD)^2}{2m_b^2} b|
\Lambda_{\mathrm{b}}(v)\rangle_{\mathrm{spin-aver.}}\,,\quad
\langle\Lambda_{\mathrm{b}}(v)|\bar b\gamma_\mu \gamma_5 b|\Lambda_{\mathrm{b}}(v)\rangle
= \left(1 + \varepsilon_b\right) \bar u\gamma_\mu\gamma_5 u\,.
\eea
We will use in the numerical evaluations below the value $K_b=0.01$, following from the
QCD sum rule calculation of \cite{Paver}. The parameter $\varepsilon_b$ will be
taken to saturate the model-independent inequality $\varepsilon_b \leq -\frac23 K_b$ 
\cite{bound}. As discussed in this reference, this inequality is saturated if certain
double insertions of the chromomagnetic operator can be neglected; QCD sum rule calculations
indicate that this is a good approximation.

  The final $\hat q^2$-integration has to be done in the limits $0\leq
\hat q^2\leq (1-\sqrt{\rho})^2$. The integration is simple for $A(\hat q^2)$
and $B(\hat q^2)$, but the integration of $C(\hat q^2)$ leads to hypergeometric 
functions because of the extra square root factor $\sqrt{\hat q^2}$.
One obtains
\bea\label{12}
\hat \Gamma_A &=& (1-8\rho+8\rho^3-\rho^4-12\rho^2\log\rho)(1-K_b)\\\label{13}
\hat \Gamma_{B_2} &=& (1+\varepsilon_b)
\left(-\frac13-30\rho^2-\frac{40}{3}\rho^3+\rho^4+\frac{32}{3}
\rho\sqrt{\rho}(1+3\rho)\right)\\
&+& K_b \left(-\frac59-40\rho-\frac{110}{3}\rho^2+\frac{64}{9}\rho^3
-\rho^4 + \frac{32}{3}\sqrt{\rho}(1+\frac{17}{3}\rho)\right)\nonumber\\
\hat \Gamma_{C_2} &=& \frac{3\pi^2}{32}\left((1+\varepsilon_b)
[(1-\sqrt{\rho})^6(1+\sqrt{\rho})
_2F_1(-\frac12,\frac52;4;z)\right.\nonumber\\\label{14}
&-&\,\left.2(1-\sqrt{\rho})^5(1+\sqrt{\rho})^2\,\,
_2F_1(-\frac12,\frac32;3;z)] \right.\\
& &\left.
+ \frac23 K_b [(1-\sqrt{\rho})^6(1+\sqrt{\rho})
_2F_1(-\frac12,\frac52;4;z)\right.\nonumber\\\nonumber
&+&\,\left.2(1-\sqrt{\rho})^5(1+\sqrt{\rho})^2\,\,
_2F_1(-\frac12,\frac32;3;z)]\right)\,.
\eea
where the hypergeometric function in Eq.~(\ref{14}) is defined as usual by
\bea
_2F_1(a,b;c;z) = \frac{\Gamma(c)}{\Gamma(b)\Gamma(c-b)}\int_0^1\mbox{d}x\,
x^{b-1}(1-x)^{c-b-1}(1-zx)^{-a}
\eea
and its argument is
\bea
z = \left(\frac{1-\sqrt{\rho}}{1+\sqrt{\rho}}\right)^2\,.
\eea

 In Table 1 we have listed the values of $\hat \Gamma_A :=A,\,
\hat \Gamma_{B_2} :=
B_2$ and $\hat \Gamma_{C_2} :=C_2$ for four different values of the mass ratio
$\rho$, including the nonperturbative $O(1/m_b^2)$ corrections.
In the brackets we list the numerical value of the ${\cal O}(1/m_b^2)$ corrections, 
which are very small and amount to 1-2\% of the tree-level contribution.
The three choices of the mass ratio squared $\rho = m_c^2/m_b^2 $ in Table 1 are taken
from the discussion
in \cite{prev}. The value $\rho=0$ is relevant for the $b\to u$ transitions
and also relevant for a comparison with the well-known results in $\mu$-decay.

  Next we calculate the reduced rate functions in system 1. It is clear 
that the unpolarized rate function $A$ is the same in both systems, and 
that the polarized rate functions $B$ and $C$ in the two systems are 
related to each other. In fact, it is not difficult to see that the 
relation between the two sets of polarized rate functions is given by
\bea\label{17}
\left(\begin{array}{c}
B_{1a}(\hat q^2,\cos\theta)\\
C_{1a}(\hat q^2,\cos\theta)\end{array}\right) =
\left(\begin{array}{cc}
\cos\theta_{12} & \sin\theta_{12}\\
-\sin\theta_{12} & \cos\theta_{12}\end{array}\right)
\left(\begin{array}{c}
B_{2a}(\hat q^2,\cos\theta)\\
C_{2a}(\hat q^2,\cos\theta)\end{array}\right)
\eea
where $\theta_{12}$ is the (polar) angle between $\vec p_X$ and $\vec
p_\ell$. This angle can be related to $\hat q^2$ and $\cos\theta$ by
\bea\label{18}
\cos\theta_{12} = (\hat q_0\cos\theta - \hat p)/(2\hat E_e)\,.
\eea
From Eq.~(\ref{17}) it is evident that the $\cos\theta$-dependence of the
polarized rate functions in system 1 is somewhat more complicated than that
in system 2.

 We do not pursue this possible line of approach any further here but
compute the rate functions $A,B_1$ and $C_1$
directly in the kinematic configuration at hand. Expressed in terms of
the hadronic and leptonic tensors $W_{\mu\nu}$,  $L_{\mu\nu}$, the
rate functions are given by
\bea\label{rate1}
A + P(B_1\cos\vartheta + C_1\sin\vartheta \cos\varphi) = 12 L_{\mu\nu}
W^{\mu\nu}\, \mbox{d}x \mbox{d}y \mbox{d}\hat q^2\,.
\eea
This differential rate is evaluated most conveniently by integrating
first over $\hat q^2$ within the limits $(0,xy)$ and subsequently over
the neutrino energy $x=(1-\rho-y, x_0)$, with 
$x_0=1-\frac{\rho}{1-y}$. Let us first list the results of these two
integrations which agree with the corresponding results obtained
in \cite{Gremm}. One has
\bea
\frac{\mbox{d}\hat\Gamma_A}{\mbox{d}y} &=& 2y^2x_0^2
(-3y+6+x_0(y-3))
+K_b\frac{4y^3}{(1-y)^2}\left(-(y-2)y-2x_0(y^2-3y+5)\right.\nonumber
\\
& &\qquad\qquad+\left.\,
x_0^2(2y^2-8y+15)-\frac23 x_0^3(y^2-5y+10)\right)\\
\frac{\mbox{d}\hat\Gamma_{B_1}}{\mbox{d}y} &=& 2(1+
\varepsilon_b)y^2x_0^2(-3y+x_0(y+1))
+ K_b\frac{4y^3}{(1-y)^2}\left(
-y^2-2x_0y(y-4)\right.\nonumber\\
& &+\left.\, x_0^2(2y^2-6y-5) - \frac23x_0^3(y^2-2y-5)\right)\\
\frac{\mbox{d}\hat\Gamma_{C_1}}{\mbox{d}y} &=&
(1+\varepsilon_b)\frac32\pi y^2\sqrt{1-y}x_0^2(2-x_0)\nonumber\\
& &+\,
K_b\frac{\pi}{8}\frac{y^3}{(1-y)^{3/2}}\left(16y + 8x_0(y-10) +
6x_0^2(-5y+20) + 5x_0^3(3y-10)\right)
\eea

While the calculation of the rate functions $A(y)$ and $B_1(y)$ is rather 
straightforward one encounters certain singular expressions
in the case of $C_1(y)$. 
In \cite{Gremm} a method for dealing with these problems has been
proposed (see also \cite{GroLi}). In the following we present an 
alternative treatment, which
offers perhaps a better perspective on the physical origin of these 
singularities. After inserting the OPE result for the hadronic tensor
into the rate formula (\ref{rate1}) one obtains
\bea
C_1(x,y,\hat q^2) = \sin\theta_{e\nu}\left(
F\delta(\hat q^2-\hat q_0^2) + 
G\delta'(\hat q^2-\hat q_0^2) + 
H\delta''(\hat q^2-\hat q_0^2) \right)\,,
\eea
with $\hat q_0^2=x+y+\rho-1$ and $\cos\theta_{e\nu}=1-2\hat q^2/(xy)$.
The integration over $\hat q^2$ can be performed straightforwardly
with the result
\bea\label{C1xy}
& &C_1(x,y) = \\
& &\theta(xy-\hat q_0^2)\theta(\hat q_0^2)
\left\{\sin\theta_{e\nu}(F-G'+H'') - \frac{2}{xy}\frac{\cos\theta_{e\nu}}
{\sin\theta_{e\nu}}(G-2H') - \frac{4}{x^2y^2}\frac{1}
{\sin^3\theta_{e\nu}}H\right\}_{\hat q^2=\hat q_0^2}\nonumber\\
& & -\,\frac{1}{(1-y)^2}[\sin\theta_{e\nu} H]_{\hat q^2=xy}
\delta'(x-x_0) + [\sin\theta_{e\nu} H]_{\hat q^2=0}
\delta'(x-1+\rho+y)\nonumber\\
& & +\,\frac{1}{1-y}\left\{ \sin\theta_{e\nu}(G-H')
- \frac{2}{xy}\frac{\cos\theta_{e\nu}}{\sin\theta_{e\nu}} H
\right\}_{\hat q^2=xy}\delta(x-x_0)\nonumber\\
& & -\,\left\{ \sin\theta_{e\nu}(G-H')
- \frac{2}{xy}\frac{\cos\theta_{e\nu}}{\sin\theta_{e\nu}} H
\right\}_{\hat q^2=0}\delta(x-1+\rho+y)\nonumber
\eea
The primes on $G,H$ denote differentiation with respect to $\hat q^2$.
The difficulty with this expression is that the last two surface terms
are divergent, since $\sin\theta_{e\nu}=0$ for $\hat q^2=0$ and 
$\hat q^2= xy$.
Therefore the result (\ref{C1xy}) is ill-defined as it stands and
must be defined in some way. We choose to do this by imposing a
cut-off $\varepsilon$ on the angle $\theta_{e\nu}$ such that 
$\theta_{e\nu}=(\varepsilon, \pi-\varepsilon)$.

Such a cut-off is implicit in any experimental extraction of the
rate function $C$. At $\theta_{e\nu}=0$ and $\pi$ the decay products
are collinear in the decay rest frame and consequently the orientation 
of the decay plane is undetermined. Therefore $C$ is practically
undefined at this kinematic point, which has to be excluded from the
analysis.

In our calculation this cutoff is implemented by integrating only over
$\hat q^2$ within the limits $\hat q^2_{min}=xy\varepsilon^2/4$,
$\hat q^2_{max}=xy(1-\varepsilon^2/4)$.  The limits on the neutrino
energy $x$ will have to be modified too, as follows
\bea\label{x'min}
x'_{min} &=& \frac{1-y-\rho}{1-y\varepsilon^2/4} = (1-y-\rho) +
y(1-\rho-y)\frac14\varepsilon^2 + O(\varepsilon^4)\\\label{x'max}
x'_{max} &=& \frac{1-y-\rho}{(1-y)+y\varepsilon^2/4} = x_0 -
\frac{y(1-\rho-y)}{(1-y)^2}\frac14\varepsilon^2 + O(\varepsilon^4)\,.
\eea
With this regularization the boundary terms in (\ref{C1xy}) give
poles of the form $1/\varepsilon$. These poles are cancelled 
after the integration over $x$ by similar singular terms arising
from the last term in the first line of (\ref{C1xy}). The $H$ term
is singular at the endpoints of the $x$ interval, as can be seen
explicitly from the expression for $\sin\theta_{e\nu}$
\bea
[\sin\theta_{e\nu}]_{\hat q^2=\hat q_0^2} =
\frac{2\sqrt{1-y}}{xy}\sqrt{(x_0-x)(x-1+\rho+y)}
\eea
Integration of the $H$ term in (\ref{C1xy}) over $x$ within
the limits (\ref{x'min}), (\ref{x'max}) will give, as mentioned,
$1/\varepsilon$-poles which exactly cancel those present in
the boundary terms. Therefore the correct result for $C_1(y)$ is
obtained, in the limit of a vanishingly small cut-off $\varepsilon
\ll 1$, by simply ignoring the surface terms and evaluating the
integral over $x$ of the first line in (\ref{C1xy}) in a minimal
subtraction prescription, which subtracts the $1/\varepsilon$ poles
arising from the modified integration limits (\ref{x'min}), (\ref{x'max}).

The $y$-integration has to be done in the limits $0\leq y\leq 1-
\rho$. Of relevance are only the spin dependent rate functions $\mbox{d}
\Gamma_{B_1}$ and $\mbox{d}\Gamma_{C_1}$ since the result of integrating 
the spin independent piece $\mbox{d}\Gamma_A$ can be checked to reproduce
the result Eq.~(\ref{12}). One obtains
\bea
\hat \Gamma_{B_1} &=& (1+\varepsilon_b-K_b)\left(-\frac13 + 4\rho
+ 12\rho^2 - \frac{44}{3}\rho^3 - \rho^4 + (12+8\rho)\rho^2\log\rho
\right)\\
\hat \Gamma_{C_1} &=& \frac{8\pi}{35}(1+\varepsilon_b-K_b)
\left(1 - 7\rho
+ 35\rho^2 + 35\rho^3 - \rho^{5/2}(56+8\rho)\right)\,.
\eea
where $\hat \Gamma_{B_1} := B_1$ and $\hat \Gamma_{C_1} := C_1$.
In Table 1 we list the numerical values for the reduced spin dependent rate
functions $\hat\Gamma_{B_1}:=B_1$ and $\hat\Gamma_{C_1}:=C_1$ for the
same four values of $\rho=m_c^2/m_b^2$. The discussion of the numerical
results will be deferred to until after the corresponding results in
system 3 are written down.

The spin dependent rate functions in system 3 can be obtained using similar
methods. The simplest way to treat this case is by exchanging 
 the electron and neutrino momenta 
in the lepton tensor $L_{\mu\nu}$. We obtain the following results
\bea
\frac{\mbox{d}\hat\Gamma_{B_3}}{\mbox{d}x} &=&
(1+\varepsilon_b)
\left( \frac{12\rho^2}{1-x} - 12x^3 - 24x^2\rho + 12x^2 - 12x\rho^2
-12\rho^2\right)\\
&+& K_b \left(\frac{8(1-y_0)^2}{1-x} - 12y_0^2 + 12y_0(2+\rho)
-20 x^3 + 12x\rho^2 + 16\rho^2 - 12 \rho - 12\right)\nonumber\\
\frac{\mbox{d}\hat \Gamma_{C_3}}{\mbox{d}x} &=&
-K_b 4\pi x^2 y_0^2\sqrt{1-x}\,.
\eea
In these expressions $y_0=1-\rho/(1-x)$ denotes the maximum value taken
by the electron energy at a given neutrino energy $x$.

The integrated angular rate functions in system 3 can be easily obtained by
integration in the $x$-interval $[0,1-\rho]$. They are given by
\bea\label{B3}
\hat \Gamma_{B_3} &=& (1+\varepsilon_b-K_b)\left(1 - 8\rho
+ 8\rho^3 - \rho^4 - 12\rho^2\log\rho\right)\\
\hat \Gamma_{C_3} &=& -K_b \frac{64\pi}{105}
\left(1-14\rho-35\rho^2+\rho^{3/2}(35+14\rho-\rho^2)\right)\,.
\eea
where again $\hat \Gamma_{B_3} := B_3$ and $\hat \Gamma_{C_3} := C_3$.

It is noteworthy that the polar analyzing power $\alpha$ in system 3 takes
the maximal value of 1 for the leading order free quark decay contribution.
This has been made manifest in Table 1 by rewriting the angular rate
$B_{3a}$ in terms of the unpolarized rate function $A$ dropping
a ${\cal O}(1/m_b^4)$ contribution. The fact that $\alpha$=1 can
be understood by rewriting the $(V-A)(V-A)$ form of the
matrix element into a $(S+P)(S-P)$ form with the help of the Fierz
transformation of the second type \cite{Piets}. One obtains in this
way
\bea
& &[\bar u(c)\gamma^\mu (1-\gamma_5)u(b)]
[\bar u(\ell^-)\gamma_\mu (1-\gamma_5)v(\bar\nu)]\\
& & =
2[\bar u(c) (1+\gamma_5)C\bar u^T(\ell^-)]
[v^T(\bar\nu)C^{-1} (1-\gamma_5)u(b)] \nonumber\\
& & = 2[\bar u(c) (1+\gamma_5) u(\ell^-)]
[\bar v(\bar\nu) (1-\gamma_5)u(b)]\nonumber
\eea
with $C$ the charge conjugation matrix. In the Fierz-rearranged form
it is clear that the $b$-spin is aligned with the spin direction
of the $\bar\nu$ which points along its
momentum direction. Thus the polar angle dependence in system 3
is given by $1+\cos\vartheta$, corresponding to a maximal polar
analyzing power in this system. Note that this argument is independent of the
value of the charm quark mass, such that the maximal value of this
asymmetry parameter is obtained for any value of the mass ratio $\rho$.
This can be seen directly from comparing (\ref{12}) and (\ref{B3}) where
the $\rho$-dependent coefficients of the free quark decay contribution
(and the $K_b$ contribution for that matter) can be seen to be equal to one another.

As mentioned above, the positivity of the decay rate for any values
of $(\vartheta, \varphi)$ requires the asymmetry parameter $B_3$
to be smaller or equal to 1. From this and the result (\ref{B3}) 
for this parameter one can obtain the constraint 
$\varepsilon_b \leq 0$ on the nonperturbative 
matrix element $\varepsilon_b$ to leading order in $\alpha_s$.
This is compatible with, although less stringent than the inequality 
$\varepsilon_b \leq -\frac23 K_b$ obtained in \cite{bound} from a 
zero recoil sum rule.

The azimuthal asymmetry vanishes at leading order in $1/m_b$ since
the polar asymmetry takes the maximal value of 1 in system 3. 
This can be understood by noting that
the positivity of the differential rate requires the combination
$|B|^2+|C|^2\cos^2\phi$ to be smaller than the unity for any $\phi$. 
This gives that $C$ must vanish if $B=1$.

When comparing the results of systems 1 and 2 one notes the equality 
of the FQD angular rate
functions $B_1=B_2$ and $C_1= - C_2$
when $\rho=0$, \,i.e. the case relevant for $b\to u$ 
transitions. This can be seen to be a consequence of the fact that the
$u$-quark and the electron are Fierz symmetric partners in the decay.
In the case of mass degeneracy as for the case discussed here, the FQD
decay 
distributions are symmetric under the exchange of the two and thus are the 
same in system 1 and 2. The minus sign in the relation $C_1= - C_2$
comes about because one is comparing system $1_a$ with system $2_b$ when
exchanging the Fierz partners.

Up to now we have given results for the fully integrated rate coefficients
$B_i$ and $C_i$ ($i=1,2,3$) for the process $b\to c+e^- + \bar\nu_e$ apart
from the rate function $A$ which is the same in all systems.
In Table 3 we give the corresponding coefficients for the processes
$\bar b\to \bar c+e^+ +\nu_e$, $c\to s+e^+ +\nu_e$ and $\bar c\to \bar s+e^- + \bar\nu_e$.
They can be obtained from the CP-invariance of the interaction and the
symmetry under the exchange $e^- \rightarrow \nu_e$ when going from
$b \rightarrow c$ to $c \rightarrow s$ in the integrated rate formula,
or by direct calculation. In the latter case we have used a general $V+xA$
interaction in order to obtain a nonvanishing result for the "$C_{3a}$"
entries. In addition, when going from $b \rightarrow c$ to $c \rightarrow s$
and $\bar c \rightarrow \bar s$ one has to replace $\rho=m_c^2/m_b^2$ by
$\rho=m_s^2/m_c^2$.

As concerns the numerical values of the polarization dependent 
contributions for $\rho=0$ we want to draw an analogy to muon decay.
For this purpose we arrange the decay products
in muon decay in the same weak isospin order as in the $b\to u$ case. One 
has
\bea
b\to 
\stackrel{\vert \hspace*{-0.5mm}\vspace*{0mm}\overline{
\mbox{\rule[0cm]{0cm}{1.4mm}}
\hspace*{8mm}}\hspace*{-0.5mm}\vspace*{0mm}\vert}
{u+\ell^-}
+\bar\nu_\ell\quad \Longleftrightarrow\quad \mu^-\to
\stackrel{\vert \hspace*{-0.5mm}\vspace*{0mm}\overline{
\mbox{\rule[0cm]{0cm}{1.5mm}}
\hspace*{8mm}}\hspace*{-0.5mm}\vspace*{0mm}\vert}
{\nu_\mu+e^-}
+\bar\nu_e
\eea
where we have drawn the braces connecting the Fierz partners for added 
emphasis. 
From comparing the two decays the value $B_1=B_2=-1/3$ in Table 1 should
be well familiar from
$\mu$-decay, when the electron mass is neglected. The result $C_1= - C_2=
\frac{8\pi}{35}$ has not been widely publicized in $\mu$-decay for the
obvious reason that its determination requires an azimuthal measurement
which cannot be done in $\mu$-decay because of the two undetected neutrinos
in the final state.

  Instead of analyzing the full ($\vartheta,\,\varphi$) two-fold angular decay
distribution one can reduce the two-fold distribution to single angle decay
distributions by doing either the $\vartheta$ integration or the $\varphi$ 
integration. One obtains
\bea
\frac{\mbox{d}\hat \Gamma}{\mbox{d}\cos\vartheta} \propto 1 + \alpha_P P\cos
\vartheta
\eea
and \bea
\frac{\mbox{d}\hat \Gamma}{\mbox{d}\varphi} \propto 1 + \gamma_P P\cos
\varphi
\eea
where, in terms of the angular coefficients $A,\,B$ and $C$, the polar
and azimuthal asymmetry 
parameters are given by $\alpha_P=B/A$ and $\gamma_P=\pi C/(4A)$,
respectively.
Note that the asymmetry parameter lie in the following intervals:
\bea
-1\leq \alpha_P\leq 1\qquad\mbox{and}\qquad -\frac{3\pi^2}{32\sqrt{2}}\leq
\gamma_P\leq
\frac{3\pi^2}{32\sqrt{2}}\qquad(\frac{3\pi^2}{32\sqrt{2}}\simeq 0.654)\,.
\eea
Table 1 also contains the numerical values of the asymmetry parameters 
$\alpha_P$ and $\gamma_P$.

\section{Spin-momentum correlations in the transversity system}

 In the transversity coordinate systems the event plane is in the
($x,\,y$)-plane. The orientation of the polarization vector $\vec P$
is specified by the polar angle $\tilde\vartheta$ and $\tilde\varphi$ as
drawn in Fig.~2. The relation between the transversity angles ($\tilde
\vartheta,\,\tilde\varphi$) and the helicity angles ($\vartheta,\,\varphi$)
can be easily seen to be given by
\bea
\cos\tilde\vartheta &=& \sin\vartheta\sin\varphi\\
\sin\tilde\vartheta\sin\tilde\varphi &=& \sin\vartheta\cos\varphi\\
\sin\tilde\vartheta\cos\tilde\varphi &=& \cos\vartheta\,.
\eea
Correspondingly one has the two-fold angular decay distribution
\bea\label{3.4}
\frac{\mbox{d}\hat \Gamma}{\mbox{d}\cos\tilde\vartheta\mbox{d}\tilde\varphi}=
\Gamma_b\left( A + P\sin\tilde\vartheta(B\cos\tilde\varphi + C\sin
\tilde\varphi)\right)\
\eea
after the internal three-fold phase space integration. We shall not
discuss the two-fold angular decay distribution in the transversity
system any further but immediately turn to
the single angle decay distributions. Again there are the two possibilities
of integrations over $\tilde\varphi$ or over $\cos\tilde\vartheta$.

Integrating Eq.~(\ref{3.4}) over $\tilde\varphi$ leads to a flat $\cos\tilde
\vartheta$-distribution while integrating over $\cos\tilde\vartheta$ one
obtains
\bea
\frac{\mbox{d}\hat \Gamma}{\mbox{d}\cos\tilde\varphi} \propto
1 + \tilde\gamma_P\cos(\tilde\varphi-\beta)
\eea
where
\bea
\tilde\gamma_P = \frac{\sqrt{B^2+C^2}}{4A}\pi
\eea
and the phase angle $\beta$ is given by
\bea
\beta = \arcsin\frac{C}{\sqrt{B^2+C^2}}\,.
\eea
In Table 2 we list the values of the asymmetry parameter $\tilde\gamma_P$
and the phase angle $\beta$ for two typical values of $\rho=m_c^2/m_b^2$
in systems 1,2 and 3.

\section{Conclusions}

We presented in this paper a study of the spin-momentum correlations in
$\Lambda_{\rm b}$ decays, including nonperturbative corrections of order
$1/m_b^2$. Numerically the nonperturbative corrections to the various
inclusive asymmetries are rather small, of the order of or smaller than 1\%.
This is very similar to the situation encountered for radiative QCD
corrections to spin-momentum correlations, which were computed in
\cite{CJKK}. They were found to be smaller than 1\% in all cases of
practical interest, due to a cancellation in the ratio of polarized and unpolarized
decay rates respectively.
One concludes therefore that the free-quark decay model can be expected to give 
accurate results for the asymmetries considered.
The results of our paper could be expected to be useful in measuring the
polarization of $\Lambda_{\rm b}$ produced in $e^+e^-$ annihilation
through their semileptonic decay products, complementing the methods
already proposed in \cite{BR,prev}.

\acknowledgements
We thank C. Diaconu for discussions on the experimental aspects of the 
$\Lambda_b$ quark polarization measurements and
M. Fischer for help with Table 3. D.P. is grateful to the
Physics Department of the Johannes Gutenberg-University, Mainz, for hospitality 
during the completion stage of this work.
This work was partially supported by the BMBF, FRG under contract
No. 06MZ865.

\newpage

\begin{center}
\begin{tabular}{|c|rrr|c|}
\hline
$\rho$   & 0.081  & 0.091  & 0.101  & 0\\
\hline
\hline
$A$      & $0.554(1-K_b)$  & $0.516(1-K_b)$  & $0.481(1-K_b)$  & $1-K_b$\\
\hline\hline
$B_{1a}$ & $-0.147(1+\varepsilon_b-K_b)$ & $-0.134(1+\varepsilon_b-K_b)$  & 
   $-0.122(1+\varepsilon_b-K_b)$ & $\displaystyle-\frac13(1+\varepsilon_b-K_b)$\\
\hline
$C_{1a}$ & $0.413(1+\varepsilon_b-K_b)$ & $0.386(1+\varepsilon_b-K_b)$  & 
   $0.360(1+\varepsilon_b-K_b)$ & $\displaystyle
                                     \frac{8\pi}{35}(1+\varepsilon_b-K_b)$\\
\hline
$\alpha_{1a}$ & $-0.265(1+\varepsilon_b)$ & $-0.259(1+\varepsilon_b)$  & 
   $-0.253(1+\varepsilon_b)$ & $\displaystyle-\frac13(1+\varepsilon_b)$\\
\hline
$\gamma_{1a}$ & $0.586(1+\varepsilon_b)$ & $0.587(1+\varepsilon_b)$  & 
   $0.589(1+\varepsilon_b)$ & $\displaystyle
                                    -\frac{2\pi^2}{35}(1+\varepsilon_b)$\\
\hline\hline
$B_{2a}$ & $-0.231(1+\varepsilon_b)$ & $-0.219(1+\varepsilon_b)$  & 
   $-0.207(1+\varepsilon_b)$ & $\displaystyle-\frac13(1+\varepsilon_b)-\frac59 K_b$\\
         & $+0.397 K_b$  & $+0.383 K_b$   & $+0.368 K_b$  &     \\
         & (+0.005)  & (+0.005)   & (+0.005)  &      \\
\hline
$C_{2a}$ & $-0.382(1+\varepsilon_b)$ & $-0.355(1+\varepsilon_b)$  & 
   $-0.329(1+\varepsilon_b)$ & $\displaystyle
         -\frac{8\pi}{35}(1+\varepsilon_b)+\frac{8\pi}{21} K_b$\\
         & $+0.446 K_b$  & $+0.405 K_b$   & $+0.369 K_b$  &         \\
         & (+0.007)  & (+0.006)   & (+0.006)  &       \\
\hline
$\alpha_{2a}$ & $-0.418$ & $-0.424$  & $-0.430$ & 
   $\displaystyle-\frac13(1+\varepsilon_b)-\frac89 K_b$\\
         & (+0.006)  & (+0.006)   & (+0.006)  &         \\
\hline
$\gamma_{2a}$ & $-0.542$ & $-0.540$  & $-0.538$ & $\displaystyle
       -\frac{2\pi^2}{35}(1+\varepsilon_b) +\frac{4\pi^2}{105} K_b$\\
         & (+0.004)  & (+0.004)   & (+0.004)  &         \\
\hline\hline
$B_{3a}$ & $A(1+\varepsilon_b)$ & $A(1+\varepsilon_b)$  & 
   $A(1+\varepsilon_b)$ & $1+\varepsilon_b - K_b$\\
\hline
$C_{3a}$ & $-0.898 K_b$  & $-0.827 K_b$   & $-0.761 K_b$  &  
   $\displaystyle -\frac{64\pi}{105}K_b$       \\
\hline
$\alpha_{3a}$ & $1+\varepsilon_b$ & $1+\varepsilon_b$  & $1+\varepsilon_b$ & 
   $1+\varepsilon_b$\\
\hline
$\gamma_{3a}$ & $-1.274 K_b$ & $-1.258 K_b$  & $-1.243 K_b$ & $\displaystyle
       -\frac{16\pi^2}{105} K_b$\\
\hline
\end{tabular}
\end{center}
\begin{quote} {\bf Table 1.} Values of spin independent reduced rates
$\Gamma_A:=A$ and spin dependent rates $\Gamma_{B_i}:=B_i$ and
$\Gamma_{C_i}:=C_i$ in the helicity systems $i=1,2,3$ for different
values of the mass ratio $\rho=m_c^2/m_b^2$. Also shown are asymmetry parameters
$\alpha_i$ and $\gamma_i$.
In brackets are shown the changes induced by the inclusion of 
${\cal O}(1/m_b^2)$ contributions corresponding to the values $K_b=0.01$ \cite{Gremm,Paver}
and $\varepsilon_b=-\frac23 K_b$ \cite{bound}.
\end{quote}

\begin{center}
\begin{tabular}{|c|r|r|}
\hline
$\rho$   & 0.091  & 0\\
\hline
\hline
$\tilde\gamma_{P_1}$      & $0.622$  & $0.622$ \\
$\beta_1$ & 1.237 & 1.136 \\
\hline
$\tilde\gamma_{P_2}$      & $0.635$  & $0.622$ \\
$\beta_2$ & $-1.018$ & $-1.136$ \\
\hline
$\tilde\gamma_{P_3}$      & $0.785$  & $0.785$ \\
$\beta_3$ & 0 & 0 \\
\hline
\end{tabular}
\end{center}
\begin{quote} {\bf Table 2.} Transversity system asymmetries in
system $i=1,2,3$ for two typical values of the quark mass ratio $\rho=m_c^2/m_b^2$.
The nonperturbative $1/m_b^2$ corrections have been neglected.
\end{quote}
\newpage

\begin{center}
\begin{tabular}{c|rrr|rrr}
\hline
\hline
 & $1a$ & $2a$ & $3a$   &  $1a$ & $2a$ & $3a$ \\
\hline
\hline
$b\to c+e^-+\bar\nu_e$   & $B_{1a}$  & $B_{2a}$  & $B_{3a}$  & $C_{1a}$ & $C_{2a}$ & $C_{3a}$ \\
$\bar b\to \bar c+e^++\nu_e$   & 
$-B_{1a}$  & $-B_{2a}$  & $-B_{3a}$  & $-C_{1a}$ & $-C_{2a}$ & $-C_{3a}$ \\
$c\to s+e^++\nu_e$   & $B_{3a}$  & $B_{2a}$  & $B_{1a}$  & $-C_{3a}$ & $-C_{2a}$ & $-C_{1a}$ \\
$\bar c\to \bar s+e^-+\bar\nu_e$   & 
$-B_{3a}$  & $-B_{2a}$  & $-B_{1a}$  & $C_{3a}$ & $C_{2a}$ & $C_{1a}$ \\
\hline
\hline
\end{tabular}
\end{center}
\begin{quote} {\bf Table 3.} 
Fully integrated angular coefficients $B_i$ and $C_i$ ($i=1,2,3$) for the processes
$b\to c+e^-+\bar\nu_e$, $\bar b\to \bar c+e^++\nu_e$, $c\to s+e^++\nu_e$ and
$\bar c\to \bar s+e^-+\bar\nu_e$ for the three coordinate systems $1a$, $2a$ and $3a$
defined in Eq.~(2.1). Results for systems $1b$, $2b$ and $3b$ can be obtained using
$B_{ib}=B_{ia}$ and $C_{ib}=-C_{ia}$.
\end{quote}

\thispagestyle{plain}
\begin{figure}[hhh]
 \begin{center}
 \mbox{\epsfig{file=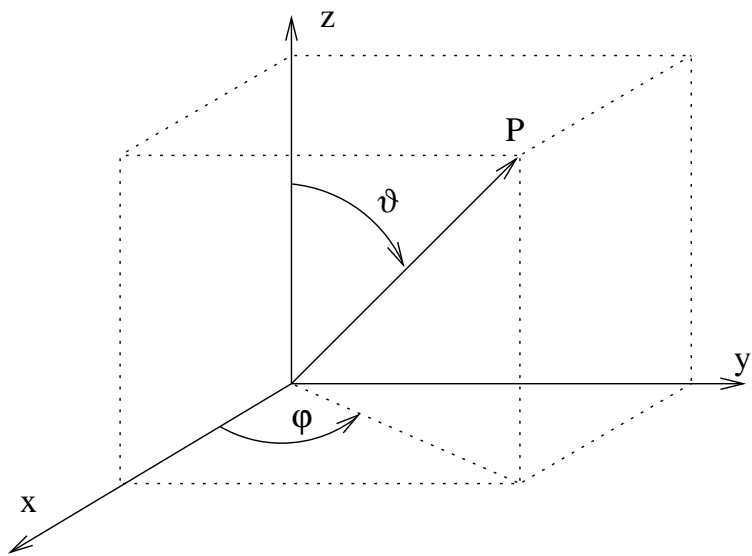,width=10cm}}
 \end{center}
 \caption{
Helicity coordinate system defining the polar angle $\vartheta$ and the azimuthal
angle $\varphi$. The decay plane is in the $(x,z)$-plane.}
\label{fig1}
\end{figure}

\thispagestyle{plain}
\begin{figure}[hhh]
 \begin{center}
 \mbox{\epsfig{file=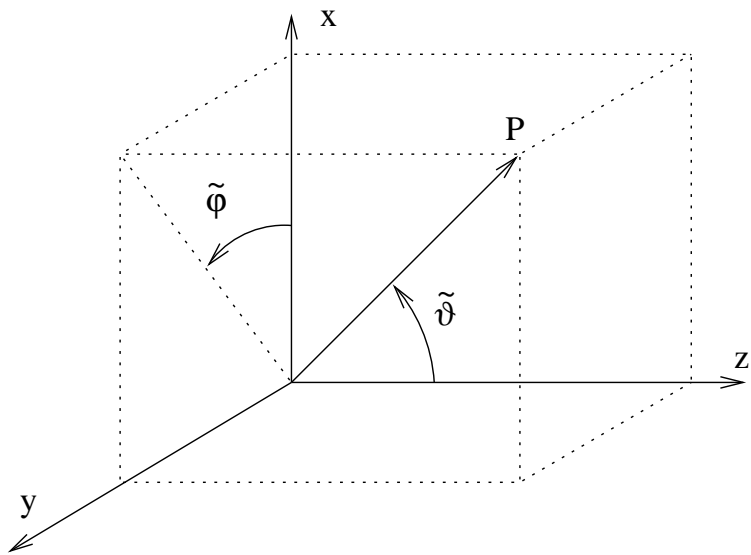,width=10cm}}
 \end{center}
 \caption{
Transversity coordinate system defining the polar angle $\tilde\vartheta$ and the azimuthal
angle $\tilde \varphi$. The decay plane is in the $(x,y)$-plane.}
\label{fig2}
\end{figure}


\begin{references}
\bibitem{PPE} The ALEPH Collaboration, 
    Phys. Lett. {\bf B365} 437 (1996).
\bibitem{tr1} J.H. K\"uhn, A. Reiter and P.M. Zerwas, Nucl. Phys.
   {\bf B272} 560 (1986).
\bibitem{OPAL} The OPAL Collaboration, CERN-EP/98-119, hep-ex/9808006.
\bibitem{BR} G. Bonvicini and L. Randall,  Phys. Rev. Lett. {\bf 73}
   392 (1994).
\bibitem{prev} C. Diaconu, J.G. K\"orner, D. Pirjol and M. Talby,
   Phys. Rev. {\bf D53} 6186 (1996).
\bibitem{JGK} J.G. K\"orner, Nucl. Phys. {\bf B} (Proc. Suppl.) {\bf 50}
   130 (1996).
\bibitem{CGG} J. Chay, H. Georgi and B. Grinstein,  Phys. Lett. {\bf B247}
   399 (1990).
\bibitem{Blok} B. Blok, L. Koyrakh, M. Shifman and A.I. Vainshtein,
   Phys. Rev. {\bf D49} 3356 (1994).
\bibitem{MW} A. Manohar and M. Wise, Phys. Rev. {\bf D49} 1310 (1994).
\bibitem{Ma} T. Mannel, Nucl. Phys. {\bf B314} 396 (1994).
\bibitem{Paver} P. Colangelo, C. A. Dominguez, G. Nardulli
   and N. Paver, Phys. Rev. {\bf D54} 4622 (1996). 
\bibitem{bound} J.G. K\"orner and D. Pirjol, Phys. Lett. {\bf B334} 399
   (1994).
\bibitem{CJKK} A. Czarnecki, M. Jezabek, J.G. K\"orner and J.H. K\"uhn,
    Phys. Rev. Lett. {\bf 73} 384 (1994).
\bibitem{CJ} A. Czarnecki and M. Jezabek,
    Nucl. Phys. {\bf B427} 3 (1994).
\bibitem{Stephan} S. Balk, J.G. K\"orner and D. Pirjol,
    Eur. Phys. J. {\bf C1} 221 (1998). Erratum, to be published.
\bibitem{Gremm} M. Gremm, G. K\"opp and L.M. Sehgal, Phys. Rev. 
   {\bf D52} 1588 (1995).
\bibitem{GroLi} Y. Grossman and Z. Ligeti,  Phys. Lett. {\bf B347} 399 (1995).
\bibitem{Piets} H. Pietschmann, {\em Weak Interactions -- Formulae,
   Results, and Derivations}, Springer-Verlag, 1983.
\bibitem{1} J.G. K\"orner, A. Pilaftsis and M.M. Tung, Z. Phys. {\bf C63}
   575 (1994). 
\bibitem{2} S. Groote, J.G. K\"orner and M.M. Tung,  Z. Phys. {\bf C74}
   615 (1997). 
\bibitem{CKPS} F. Close, J.G. K\"orner, R.J.N. Phillips and D.J. Summers,
   J. Phys. {\bf G18} 1716 (1992).
\end{references}
\end{document}